# Modeling Bimodal Discrete Data Using Conway-Maxwell-Poisson Mixture Models

Pragya Sur[a], Galit Shmueli[b], Smarajit Bose[a], Paromita Dubey[a]

[a]Indian Statistical Institute, Kolkata 700108, India
[b]Indian School of Business, Hyderabad 500032, India

## Abstract

Bimodal truncated count distributions are frequently observed in aggregate survey data and in user ratings when respondents are mixed in their opinion. They also arise in censored count data, where the highest category might create an additional mode. Modeling bimodal behavior in discrete data is useful for various purposes, from comparing shapes of different samples (or survey questions) to predicting future ratings by new raters. The Poisson distribution is the most common distribution for fitting count data and can be modified to achieve mixtures of truncated Poisson distributions. However, it is suitable only for modeling equi-dispersed distributions and is limited in its ability to capture bimodality. The Conway-Maxwell-Poisson (CMP) distribution is a two-parameter generalization of the Poisson distribution that allows for over- and under-dispersion. In this work, we propose a mixture of CMPs for capturing a wide range of truncated discrete data, which can exhibit unimodal and bimodal behavior. We present methods for estimating the parameters of a mixture of two CMP distributions using an EM approach. Our approach introduces a special two-step optimization within the M step to estimate multiple parameters. We examine computational and theoretical issues. The methods are illustrated for modeling ordered rating data as well as truncated count data, using simulated and real examples.

**Keywords**: censored data, count data, EM algorithm, Likert scale, surveys

# 1 Introduction and Motivation

Discrete data arise in many fields, including transportation, marketing, healthcare, biology, psychology, public policy, and more. Two particularly common types of discrete data are ordered ratings (or rankings) and counts. This paper is motivated by the need for a flexible distribution for modeling discrete data that arise in truncated environments, and in particular, where the empirical distributions exhibit bimodal behavior. One example is aggregate counts of responses to Likert scale questions or ratings such as online ratings of movies and hotels, typically on a scale of one to five stars. Another context where bimodal truncated discrete behavior is observed is when only a censored version of count data is available. For example, when the data provider combines the highest count values into a single "larger or equal to" bin, the result is often another mode at the last bin.

Real data in the above contexts can take a wide range of shapes, from symmetric to left- or right-skewed and from unimodal to bimodal. Peaks and dips can occur at the extremes of the scale, in the middle, etc. Data arising from ratings or Likert scale[1] questions exhibit bimodality when the respondents have mixed opinions. For example, respondents might have been asked to rate a certain product on a ten-point scale. If some respondents like the item considerably and others do not, we would find two modes in the resulting data, and the location of the modes would depend on the extent of the likes and dislikes. In online ratings, sometimes the owners of the rated product/service illegally enter ratings, thereby contributing to overly "good" ratings, while other users might report very "bad" ratings. This behavior would again result in bimodality.

In addition to bimodality, data from different groups of respondents might be under-dispersed or over-dispersed, due to various causes. For example, dependence between responders' answers can cause over-dispersion.

---

[1] An example of a typical 5-point Likert scale is: strongly disagree, disagree, neutral (undecided), agree, strongly agree

The most commonly used distribution for modeling count data is the Poisson distribution. One of the major features of the Poisson distribution is that the mean and variance of the random variable are equal. However, data often exhibit over- or under-dispersion. In such cases, the Poisson distribution often does not provide good approximations. For over-dispersed data, the negative Binomial model is a popular choice (Hilbe, 2010). Other over-dispersion models include Poisson mixtures (McLachlan, 1997). However, these models are not suitable for under-dispersion. A flexible alternative that captures both over- and under-dispersion is the Conway-Maxwell-Poisson (CMP) distribution. The CMP is a two-parameter generalization of the Poisson distribution which also includes the Bernoulli and geometric distributions as special cases (Shmueli et al., 2005). The CMP distribution has been used in a variety of count-data applications and has been extended methodologically in various directions (see a survey of CMP-based methods and applications in Sellers et al., 2012).

In the context of bimodal discrete data, and for capturing a wide range of observed aggregate behavior, we therefore propose and evaluate the use of a mixture of two CMP distributions. We find that a mixture of Poisson distributions is often insufficient for adequately capturing many bimodal distribution shapes. Consider, for example, the situation of responses with a U-shape with one peak at a low rating (say, 1), followed by a steep decline, a deep valley, and then a sudden peak at a high rating (say,9). A mixture of two Poisson distributions will likely be inadequate due to the steep decline after 1 and sudden rise near 9. Such data might arise from a mixture of two under-dispersed distributions. There might be other situations where the data can be conceived of as a mixture of two over-dispersed distributions or an over-dispersed and an under-dispersed distribution. Under such setups, mixtures of two CMP distributions are likely to better fit the data than mixtures of two Poisson distributions. While the CMP distribution has been the basis for various models, to the best of our knowledge, it was not extended to mixtures.

A model for approximating truncated discrete bimodal data is useful for various goals. By approximating, we refer to the ability to estimate the locations and magnitudes of the peaks and dips of the distribution. One application is prediction, where the purpose is to predict the magnitude of the outcome for new observations (such as in online ratings). Another is to try and distinguish between two underlying groups (such as between fraudulent self-rating providers and legitimate raters).

We are interested both in the frequency of a given value as well as in the value itself. In the case of "popular" values, we use the term "peak" to refer to the magnitude and "mode" to refer to the location of the peak. In the case of "unpopular" values, we use the term "dip" to refer to the magnitude, and coin the term "lode" to refer to the location of the dip. In bimodal data, we expect to see two peaks and one, two, or three dips. We denote these by *mode$_1$*, *mode$_2$*, *lode$_1$*, *lode$_2$*, *lode$_3$*, where *mode$_1$* and *lode$_1$* are the left-most (or top-most) mode and lode on a vertical (horizontal) bar chart, respectively.

In the following, we introduce three real data examples to illustrate the motivation for our proposed methodology.

## 1.1 Example 1: Survey question in market research

Market research companies conduct surveys to better understand consumers' preferences. Such surveys often include 5-point Likert scales. Figure 1 shows the empirical distribution of 199 respondents to a question regarding an ice-cream product. The question asks about the presence of ice pieces in a particular ice-cream product. We can see a bimodal distribution, with a peak at "1" and another at "3". Using our terminology: mode1=1, mode2=3, lode1=2, lode$_2$=5.

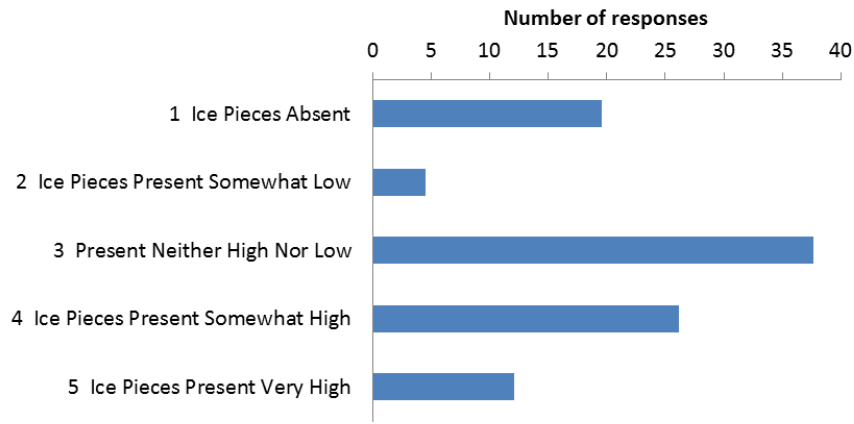

Figure 1: Distribution of responses to question on ice-cream product

## 1.2 Example 2: Censored data

The Heritage Provider Network is, a healthcare provider, recently launched a $3,000,000 contest with the following goal: "Identify patients who will be admitted to a hospital within the next year, using historical claims data."[2] While the contest is much broader, for simplicity we look at one of the main outcome variables, which is the distribution of the number of days spent in the hospital (for claims received in a two-year period)[3]. The censoring at 15 days of hospitalization creates a second mode in the data, as can be seen in Figure 2. In this example, $mode_1=1$, $mode_2=15+$, $lode1=14$.

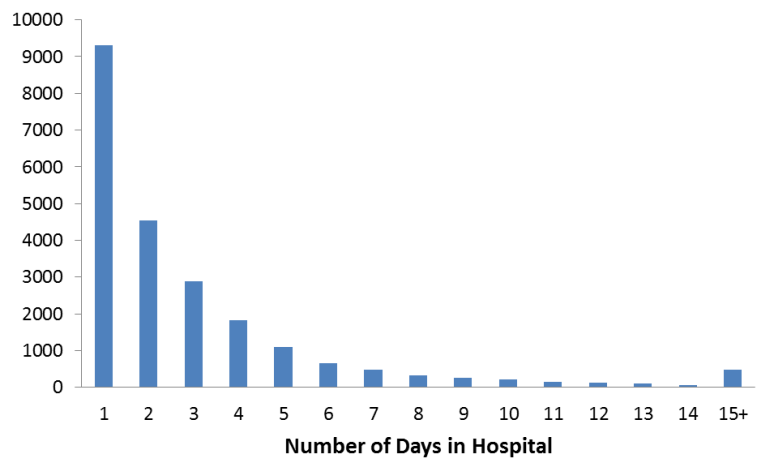

Figure 2: Distribution of numbers of days at the hospital. Data reported in censored form

## 1.3 Example 3: Online ratings

Many websites rely on user ratings for different products or services, and a "5 star" rating system is common. Amazon.com, netflix.com, tripadvisor.com are just a few examples of such websites. To illustrate such a scenario, Figure 3 shows the ratings for a hotel in Bhutan as displayed on the popular travel website tripadvisor.com (the data were recorded on May 24, 2012 and can change as more ratings are added by users). In this example, we see bimodal behavior that reflects mixed reviews. Some responders have an "excellent" or "very good" impression of the hotel while a few report a "terrible" experience. Here, $mode_1=Excellent$, $mode_2=Terrible$, $lode_1=Poor$.

---

[2] http://www.heritagehealthprize.com
[3] We excluded zero counts which represent patients who were not admitted at all. The latter consist of nearly 125,000 records.

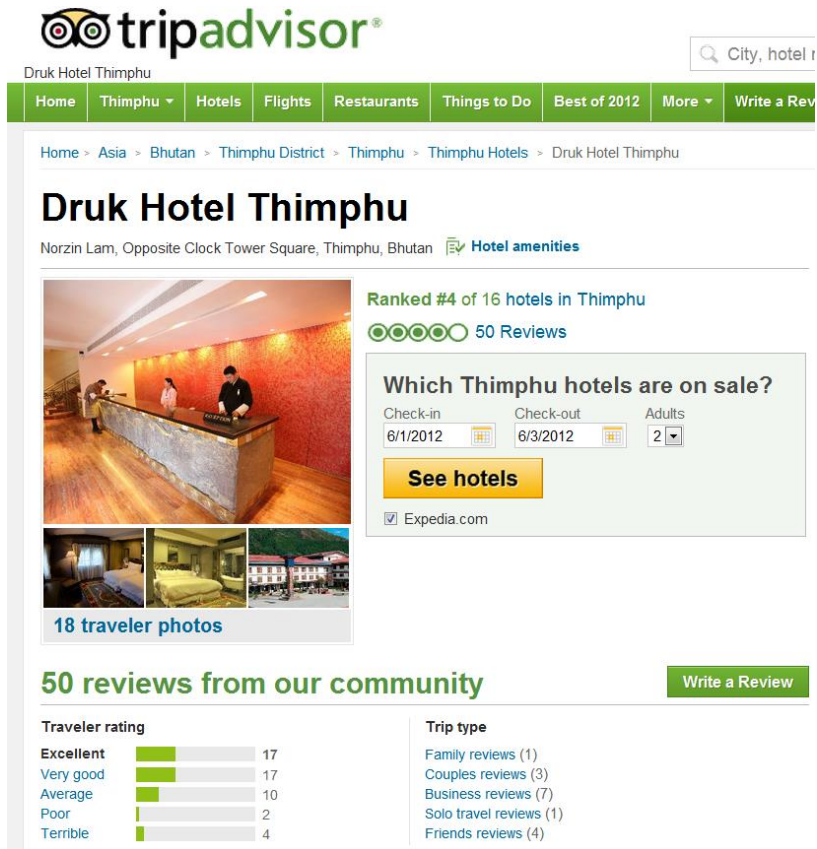

Figure 3: Distribution of user ratings for a hotel (5-point scale)

The remainder of the paper is organized as follows: In Section 2 we introduce a mixture of truncated CMP distributions for capturing bimodality, and describe the EM algorithm for estimating the five CMP mixture parameters and computational considerations. We also discuss measures for comparing model performance. Section 3 illustrates our proposed methodology by applying it to simulated data, and Section 4 applies it to the three real data examples. We conclude the paper with a discussion and future directions in Section 5.

## 2 A Mixture of Truncated CMP Distributions

### 2.1 The CMP Distribution

The Conway-Maxwell-Poisson (CMP) distribution is a generalization of the Poisson distribution obtained by introducing an additional parameter ν, which can take any non-negative real value, and accounts for the cases of over and under dispersion in the data. The distribution was briefly introduced by Conway and Maxwell in 1962 for modeling queuing systems with state-dependent service rates. Non-Poisson data sets are commonly observed these days. Over-dispersion is often found in sales data, motor vehicle crashes counts, etc. Under-dispersion is often found in data on word length, airfreight breakages, etc. (see Sellers et al., 2012 for a survey of applications). The statistical properties of the CMP distribution, as well as methods for estimating its parameters were established by Shmueli et al. (2005). Various CMP-based models have since been published, including CMP regression models (classic and Bayesian approaches), cure-rate models, and more. The various methodological developments take advantage of the flexibility of the CMP distribution in capturing under- and over-dispersion, and applications have shown its usefulness in such cases. However, to the best of our knowledge, there has not been an attempt to fit bimodal count distributions using the CMP. The use of CMP mixtures is advantageous compared to Poisson mixtures, as it allows the combination of data with different dispersion levels with a resulting bimodal distribution.

If X is a random variable from a CMP distribution with parameters $\lambda$ and $\nu$, its distribution is given by

$$P(X = x) = \frac{\lambda^x}{x!^\nu} \cdot \frac{1}{\sum_{j=0}^{\infty} \frac{\lambda^j}{j!^\nu}}, for \ x = 0,1,2,\dots \ \lambda > 0, \nu \geq 0 \qquad (1)$$

It is common to denote the normalizing factor by $Z(\lambda, \nu) = \sum_{j=0}^{\infty} \frac{\lambda^j}{j!^\nu}$. The common features of this distribution are:

The ratio of successive probabilities is non-linear in x unlike that for the Poisson distribution.

$$\frac{P(X = x - 1)}{P(X = x)} = \frac{x^\nu}{\lambda}$$

In case of the Poisson distribution ($\nu$ = 1) the above quantity becomes linear (x/$\lambda$).

If v < 1, successive ratios decrease at a slower rate compared to the Poisson distribution giving rise to a longer tail. This corresponds to the case of over-dispersion. The reverse occurs for the case of under-dispersion.

This distribution is a generalization of a number of discrete distributions:

- For v=0 and λ < 1, this is a geometric distribution with parameter (1-λ).
- For v=1, this is the Poisson distribution with parameter λ.
- For v → ∞, this is a Bernoulli distribution with parameter $\lambda/(1+\lambda)$.

The CMP distribution is a member of the exponential family and $(\sum_{i=1}^{n} x_i, \sum_{i=1}^{n} \log(x_i!))$ is sufficient for (λ, v).

We modify the CMP distribution to the truncated scenario considered in this paper. For data in the range t, t+1, t+2,…, T, we truncate values below t and above T. For example, for data from a 10-point Likert scale, the truncated CMP pmf is given by:

$$P(X = x) = \frac{\lambda^x}{x!^v} \cdot \frac{1}{\sum_{j=1}^{10} \frac{\lambda^j}{j!^v}}, \quad x = 1,2, \ldots, 10 \ ; \ \lambda > 0, v \geq 0 \qquad (2)$$

## 2.2 CMP Mixtures

The principal objective of this paper is to model bimodality in count data. Since both the Poisson and CMP can only capture unimodal distributions, for capturing bimodality we resort to mixtures. The standard technique for fitting a mixture distribution is to employ the Expectation-Maximization (EM) algorithm (Dempster et al., 1977). For example, in case of Poisson mixtures, one assumes that the underlying distribution is a mixture of two Poisson component distributions with unknown parameters while the mixing parameter *p* is also unknown. Further it is also assumed that there is a hidden variable with a Bernoulli(*p*) distribution, which determines from which component the data is coming from. Starting with some initial values of the unknown parameters, in the first step (E-step) of the algorithm, the conditional expectation of the missing hidden variables are calculated. Then, in the second step (M-step), parameters are estimated by maximizing the full likelihood (where the values of the hidden variables are replaced with the expected values calculated in the E-step). Using these new estimates, the E-step is repeated, and iteratively both steps are continued until convergence.

Let X be a random variable assumed to have arisen from a mixture of $CMP(\lambda_1, \nu_1)$ and $CMP(\lambda_2, \nu_2)$ with probability $p$ of being generated from the first CMP distribution. We also assume that each CMP is truncated to the interval [1, 2,…, T].

Let $f_1(x)$ and $f_2(x)$ denote the pmfs of the two CMP distributions respectively. Then the pmf of X is given by

$$f(x) = pf_1(x) + (1-p)f_2(x) \quad for\ x = 1, 2, \ldots, T \qquad (3)$$

If $X_1, X_2, \ldots, X_n$ are iid random variables from the above mixture of two CMP distributions, their joint likelihood function is given by

$$L' = \prod_{i=1}^{n} f(x_i) = \prod_{i=1}^{n} \{pf_1(x_i) + (1-p)f_2(x_i)\}$$

$$logL' = \sum_{i=1}^{n} log\{pf_1(x_i) + (1-p)f_2(x_i)\}$$

$$= \sum_{i=1}^{n} log\left\{ p \cdot \frac{\lambda_1^{x_i}}{x_i!^{\nu_1}} \cdot \frac{1}{\sum_{j=1}^{T} \frac{\lambda_1^j}{j!^{\nu_1}}} + (1-p) \cdot \frac{\lambda_2^{x_i}}{x_i!^{\nu_2}} \cdot \frac{1}{\sum_{j=1}^{T} \frac{\lambda_2^j}{j!^{\nu_2}}} \right\} \qquad (4)$$

We would like to find the estimates $(\hat{p}, \widehat{\lambda_1}, \widehat{\nu_1}, \widehat{\lambda_2}, \widehat{\nu_2})$ by maximizing the likelihood function. However, due to the non-linear structure of the likelihood function, differentiating it with respect to each of the parameters and equating the partial derivatives to zero does not yield a closed form solution for any of the parameters. We therefore adapt an alternative procedure for representing the likelihood function.

Define a new set of random variables $Y_i$ as follows:

$Y_i = 1 \text{ if } X_i \sim CMP(\lambda_1, v_1)$

$\phantom{Y_i} = 0 \text{ if } X_i \sim CMP(\lambda_2, v_2)$

Then the likelihood and log-likelihood functions can be written as

$$L = \prod_{i=1}^{n} \left\{ (pf_1(x_i))^{y_i} ((1-p)f_2(x_i))^{(1-y_i)} \right\}$$

$$l = \log L = \sum_{i=1}^{n} y_i \{\log(p) + \log f_1(x_i)\} + \sum_{i=1}^{n}(1-y_i)\{\log(1-p) + \log f_2(x_i)\} \quad (5)$$

From here we get a closed form solution for $\hat{p}$ by differentiation:

$$\frac{\delta l}{\delta p} = 0 \Rightarrow \hat{p} = \frac{\sum_{i=1}^{n} y_i}{n}$$

The problem lies in the fact that the $y_i$'s are unknown. We therefore use the EM algorithm technique.

### E Step

Here we replace the $y_i$'s with their conditional expected value

$$\tilde{Y}_i := E(Y_i | X_i = x_i) = \frac{pf_1(x_i)}{pf_1(x_i) + (1-p)f_2(x_i)}. \quad (6)$$

### M Step

Thus, by replacing the unobserved $y_i$'s in the E-step, we get

$$\hat{p} = \frac{\sum_{i=1}^{n} \tilde{y}_i}{n}. \quad (7)$$

For the other parameters, none of the equations

$$\frac{\delta l}{\delta \lambda_1} = 0, \quad \frac{\delta l}{\delta v_1} = 0, \quad \frac{\delta l}{\delta \lambda_2} = 0, \quad \frac{\delta l}{\delta v_2} = 0$$

yield closed form solutions. We propose an iterative technique for obtaining the remaining estimates by maximizing *L*.

Because an estimate of *p* is easy to obtain, we must only maximize the likelihood based on the remaining four parameters and then iterate. In particular: Plug in $\hat{p}$ in the likelihood function *L*. Then *L* becomes a function of $\lambda_1, v_1, \lambda_2, v_2$.

The idea is to use the grid search technique to maximize *L*. In this technique, we divide the parameter space into a grid, evaluate the function at each grid point, and find the grid point where the maximum is obtained. Then, a neighborhood of this grid point is further divided into finer areas and the same procedure is repeated until convergence. We continue until the grid spacing is sufficiently small. This approach is expected to yield the correct solution as CMP distribution is a member of the exponential family. Wu (1983) established the convergence of EM for the exponential family when the likelihood turns out to be unimodal.

Since we have four parameters to estimate, carrying out a grid search for all of them simultaneously is computationally infeasible. We therefore propose a two-step algorithm. First, we fix any two of the parameters at some initial value and carry out a grid search for the remaining two. Then, fixing the values of the estimated parameters in the first step, we carry out a grid search for the remaining two.

One question is which two parameters should one fix initially. From simulation studies, we observed that fixing the $\lambda's$ and obtaining $\hat{v}'s$ and then carrying out a grid search for estimating the λ's reduces the run time of the algorithm.

## 2.3 Model Estimation

To avoid identifiability issues, if the empirical distribution exhibits a single peak, *p* is set to zero and a single CMP is estimated using ordinary maximum likelihood estimation (as in Shmueli et al., 2005) with adjustment for the truncation. Otherwise, if the empirical distribution shows two peaks, we execute the following steps:

### Initialization

Fit a Poisson mixture. If the resulting estimates of $\lambda_1$, $\lambda_2$ are sufficiently different, use these three estimates as the initial values for *p*, $\lambda_1$, and $\lambda_2$ and set the initial $\nu_1 = \nu_2 = 1$.

If the estimated Poisson mixture fails to identify a mixture of different distributions, that is, when $\lambda_1$ and $\lambda_2$ are very close, then use the estimated *p* as the initial mixing probability, but initialize $\lambda$'s by fixing them at the two peaks of the empirical distribution and set the initial $\nu_1 = \nu_2 = 1$.

Alternatively, initialize $\lambda$'s by fixing them at the two peaks of the empirical distribution but initialize $\nu$'s by using the ratio between frequencies at the peak and its neighbor(s).

### Iterations

After fixing the five parameters at initial values, the two-step optimization follows the following sequence:

For a given *p*,

- Optimize the likelihood for $\nu$'s, fixing *p*, $\lambda_1$ and $\lambda_2$ using a grid search.
- The optimal $\nu_1$, $\nu_2$ are then fixed (along with *p*). A grid search finds the optimal $\lambda_1$, $\lambda_2$
- Repeat steps 1 and 2 until some convergence stopping rule is reached
- Once the $\lambda$'s and $\nu$'s are estimated, go back to estimate *p*.
- Finally, the E step and M step are run until convergence.

### Empirical observations for improving and speeding up the convergence:

- Split the grid search for $\nu$'s into three areas: [0,0.7], (0.7,1], >1

- Grid ranges and resolution can be changed over different iterations.
- Even when the initial values are not based on the Poisson mixture, the likelihood of the Poisson mixture must be retained and used as a final benchmark, to assure that the chosen CMP mixture is not inferior to a Poisson mixture. In all our experiments, the alternative initialization described above yielded better solutions.

## 2.4 Model Evaluation and Selection

We focus on two types of goals: a purely descriptive goal, where we are looking for an approximating distribution that captures the empirical distribution, and a predictive goal where we are interested in the accuracy of predicting new observations.

In the context of bimodal ratings and truncated count data, it is desirable that the fitted distribution should capture the modes, lodes and shape of the data, as well as have a close match between the observed and expected counts. Because the data are limited to a relatively small range of values, we can examine the complete actual and fitted frequency tables. It is practical and useful to start with a **visual evaluation** of the fitted distribution(s) overlaid on the empirical bar chart. The visual evaluation can be used to compare different models and to evaluate the fit in different areas of the distribution, rather than relying on a single overall measure. Performance is therefore a matter of capturing the *shape* of the empirical distribution. One example is in surveys, where it is often of interest to compare the distributions of answers to different questions to one another, or to an aggregate of a few questions.

In the bimodal context, it is typically important to properly capture the mode(s) and lode(s). The locations of the popular and unpopular values and their extremeness within the range of values can be of importance, for instance, in ratings.

For these reasons, rather than rely on an overall "average" measure of fit, such as likelihood-based metrics, we focus on reporting the modes and lodes as well as looking at the magnitudes of the deviation at peaks and dips. We report AIC statistics only for the purposes of illustrating their uninformativeness in this context. In applications where the costs of misidentifying a mode or lode can be elicited, a cost-based measure can be computed.

## 3  Application to Simulated Data

To illustrate and evaluate our CMP mixture approach and to compare it to simpler Poisson mixtures, we simulated bimodal discrete data over a truncated region, similar to the examples of real data shown in Section 1.

### 3.1  Example 1: Bimodal distribution on 10-point scale

We start by simulating data from a mixture of two CMP distributions on a 10-point scale, one under-dispersed ($\lambda_1=1$, $\nu_1=3$) and the other over-dispersed ($\lambda_2=8$, $\nu_2=0.7$), with mixing parameter $p$=0.3. Figure 4 shows the empirical distribution for 100 observations simulated from this distribution. We see a mode at 1 and another at 10. We first fit a Poisson mixture, resulting in the fit shown in Table 1 and Figure 4. As can be seen, the Poisson mixture properly captures the two modes, but their peak magnitudes are incorrectly flipped (thereby identifying the highest peak at 1); it also does not capture the single lode at 3, but rather estimates a longer dip throughout 3,4,5. Finally, the estimated overall U-shape is also distorted. Note that the three estimated parameters ($\lambda_1, \lambda_2$ and $p$) are quite close to the generating ones, yet the resulting fit is poor.

We then fit a CMP mixture using the algorithm described in Section 4. The results are shown in Table 1 and Figure 4. The fit appears satisfactory in terms of correctly capturing the two modes and single load as well as the magnitudes of the peaks and dip. Note that the AIC statistic is very close to that from the Poisson mixture, yet the two models are visibly very different in terms of capturing modes, lodes, magnitudes and the overall shape.

Although the good fit of the CMP mixture might not be surprising (because the data were generated from a CMP mixture), it is reassuring that the algorithm converges to a solution with good fit. We also note that the estimated parameters are close to the generating parameters. Finally, we note that the runtime was about a minute.

Table 1: Simulated 10-point data (n=100) and expected counts from Poisson and CMP mixtures

| Value | Simulated Data | Poisson Mixture | CMP Mixture |
|---|---|---|---|
| 1 | 22 | 36 | 22 |
| 2 | 2 | 7 | 2 |
| 3 | 0 | 1 | 0 |
| 4 | 1 | 1 | 1 |
| 5 | 1 | 1 | 2 |
| 6 | 4 | 3 | 4 |
| 7 | 7 | 6 | 7 |
| 8 | 15 | 10 | 13 |
| 9 | 22 | 15 | 20 |
| 10 | 26 | 20 | 29 |
| **Estimates** | | | |
| $p$ | 0.3 | 0.32 | 0.24 |
| $\lambda_1, \lambda_2$ | 1,8 | 0.41, 13.58 | 1.13, 9.00 |
| $\nu_1, \nu_2$ | 3, 0.7 | | 3.75, 0.8 |
| **First Mode** | 1 | 1 | 1 |
| **Second Mode** | 10 | 10 | 10 |
| **First Lode** | 3 | 3,4,5 | 3 |
| **Second Lode** | - | - | - |
| **Third Lode** | - | - | - |
| **AIC** | | 370.6 | **370.0** |

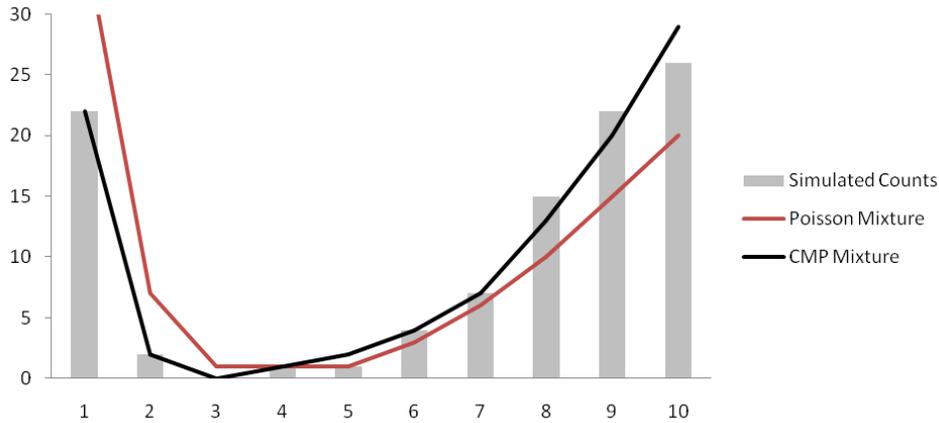

Figure 4: Fit of estimated Poisson mixture (p=0.3221, $\lambda_1$=0.4094, $\lambda_2$=13.5844) and CMP mixture (p=0.24, $\lambda_1$=1.1, $\nu_1$=3.75, $\lambda_2$=9, $\nu_2$=0.8)

## Identifiability

Identifiability can be a challenge in some cases and a blessing in other cases. When the goal is to capture the underlying dispersion level, then identifiability is obviously a challenge. However, for descriptive or predictive goals, the ability to capture the empirical distribution with more than one model allows for flexibility in choosing models based on other important considerations such as computational speed or predictive accuracy.

Exploring the likelihood function, which is quite flat in the area of the maximum, we observe an identifiability issue. In particular, we find multiple parameter combinations that yield very similar results in terms of the estimated distribution. For instance, in our above example, the estimated CMP mixture is of one under-dispersed CMP ($\lambda_1$=1.13, $\nu_1$=3.75) and one over-dispersed CMP ($\lambda_2$=9, $\nu_2$=0.8) with mixing parameter $p$=0.24. By replacing only the over-dispersed CMP with the under-dispersed CMP($\lambda_2$=25, $\nu_2$=1.27), we obtain a nearly identical fit, as shown in Figure 5 andTable2("CMP Mixture 2"). Mixture 2 is inferior to Mixture 1 only in terms of detecting lode 1 (indicating a lode at 1-2), but otherwise very similar. Another similar fit can be achieved by slightly modifying the two parameters to $\lambda_2$=30, $\nu_2$=1.36 ("CMP Mixture 3"). In other words, we can achieve similar results by combining different dispersion levels. In this example, we are able to achieve similar results by combining an over- and an under-dispersed CMP and by combining two under-dispersed CMPs.

Table 2: Three CMP mixtures fitted to the same data, with very similar fit

| Value | Counts | CMP Mixture 1($\lambda_2$=9, $\nu_2$=0.8) | CMP Mixture 2($\lambda_2$=25, $\nu_2$=1.27) | CMP Mixture 3($\lambda_2$=30, $\nu_2$=1.36) |
|---|---|---|---|---|
| 1 | 22 | 22 | 21 | 22 |
| 2 | 2 | 2 | 3 | 2 |
| 3 | 0 | 0 | 0 | 0 |
| 4 | 1 | 1 | 0 | 0 |
| 5 | 1 | 2 | 1 | 1 |
| 6 | 4 | 4 | 4 | 4 |
| 7 | 7 | 8 | 8 | 8 |
| 8 | 15 | 13 | 14 | 14 |
| 9 | 22 | 20 | 21 | 21 |
| 10 | 26 | 28 | 28 | 28 |

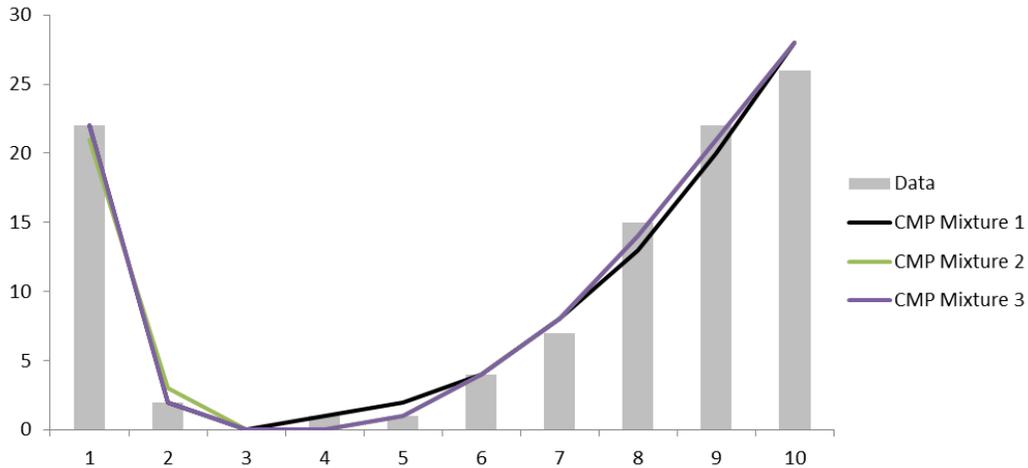

Figure 5: Three different CMP Mixture models that achieve nearly identical fit. Model 1 = CMP from Figure 6. Models 2 and 3 are mixtures of two under-dispersed CMPs

To illustrate this issue further, we also show in Figure 6 the contours of the log-likelihood functions (as functions of $\nu_1$ and $\nu_2$) for three different fixed sets of values of $p$, $\lambda_1$ and $\lambda_2$. These plots correspond to the parameter combinations given in Table 3. In the plots, only the region where the log-likelihood function nears its peak is shown. It is quite evident from the plots that the peaks of the functions achieve very similar values. Therefore the algorithm may converge to any of these parameter combinations, and we have already observed that the fits are very similar as well.

These plots also highlight the challenges of maximizing the likelihood in this situation. The solution is highly dependent on the initial values. The estimated value of the parameter $\nu_2$ depends on the estimated parameter of $\lambda_2$, and the former increases with the latter. In the process, the estimated second CMP distribution moves from being over-dispersed to even under-dispersed.

Further, in each of the plots, it can be seen that the peaks are very sharp. Therefore it is quite difficult for the algorithm to locate them. In the grid-search, the algorithm has to use very fine grids to successfully capture them. Peaks may not be visible in higher resolution. Thus the computational cost of the algorithm increases substantially.

## 3.2  Example 2: Bimodal distribution on 5-point scale

With a smaller scale, it is more difficult to detect mixture distributions, especially when the peaks become closer. In the following, we simulated data from a mixture of two CMP distributions on a 5-point scale, one under-dispersed ($\lambda_1=1$, $\nu_1=1.5$) and the other over-dispersed ($\lambda_2=5$, $\nu_2=0.7$), with mixing parameter $p$=0.3. Figure 7 and Table 3 show the empirical distribution for 100 observations simulated from this distribution. We see a first mode at 1, another mode at 5, and a dip at 2.

From Table 3, it is evident that the fitted CMP mixture correctly identifies the slight U-shape of the empirical distribution, with peaks at 1 and at 5 and a dip at 2. In contrast, while the Poisson mixture identifies the two modes, the peak heights are flipped (mode 1 is identified as the highest peak). In addition, the Poisson mixture does not identify the lode correctly. Once again, note that the AIC statistics are uninformative for comparing the models in terms of peaks, dips, and shape.

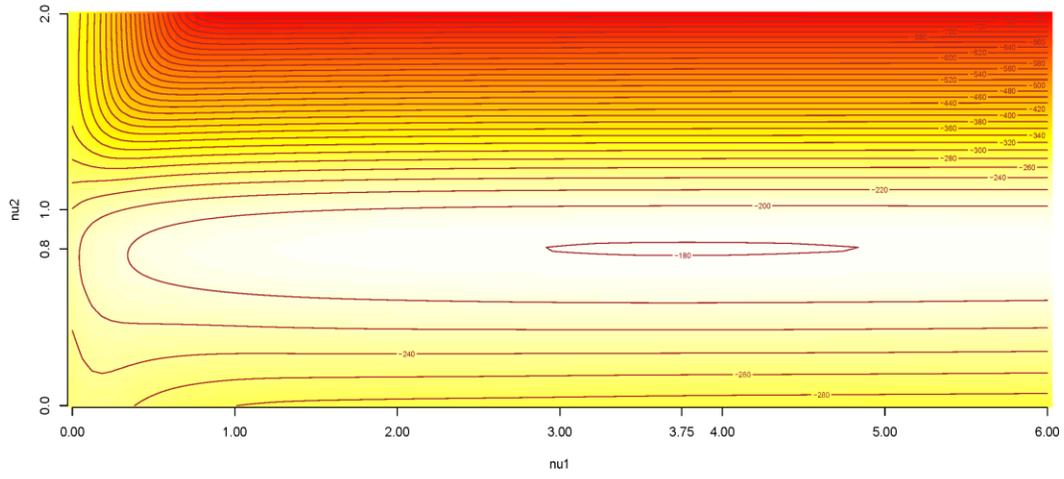

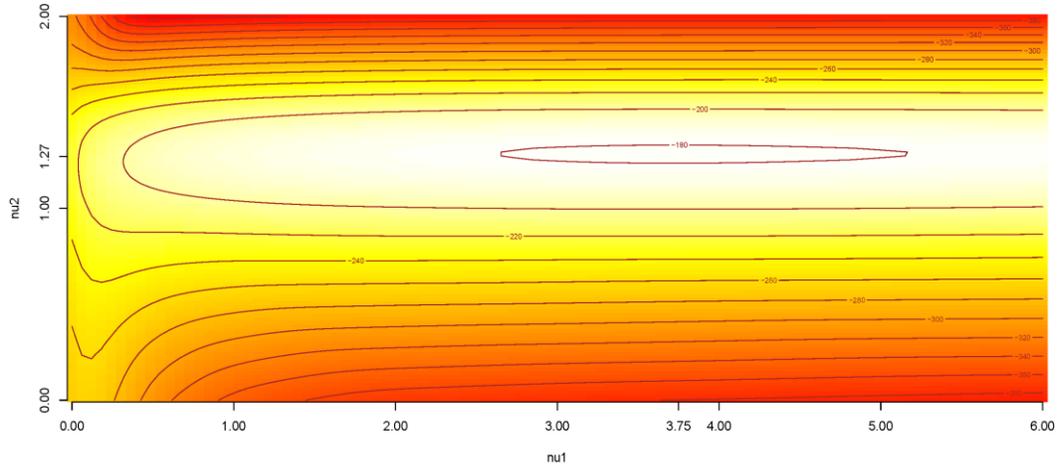

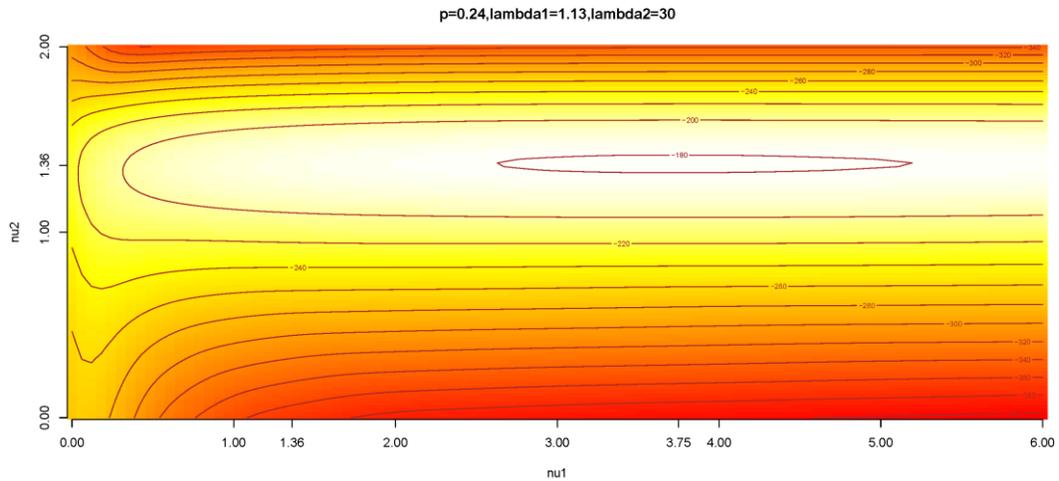

Figure 6: Contour plots of the log-likelihood for three different parameter combinations

Table 3: Simulated 5-point count data and expected counts from Poisson and CMP mixtures

| Value | Simulated Counts | Poisson Mixture | CMP Mixture |
|---|---|---|---|
| 1 | 25 | 29 | 25 |
| 2 | 8 | 14 | 9 |
| 3 | 15 | 13 | 11 |
| 4 | 16 | 19 | 21 |
| 5 | 36 | 25 | 34 |
| **Estimates** | | | |
| P | 0.3 | 0.2956 | 0.28 |
| $\lambda_1, \lambda_2$ | 1, 5 | 0.72, 6.82 | 0.99, 4.99 |
| $\nu_1, \nu_2$ | 1.5, 0.7 | | 2.3, 0.7 |
| **First Mode** | 1 | 1 | 1 |
| **Second Mode** | 5 | 5 | 5 |
| **First Lode** | 2 | 3 | 2 |
| **Second Lode** | – | - | - |
| **Third Lode** | – | - | - |
| **AIC** | | 349.02 | 338.83 |

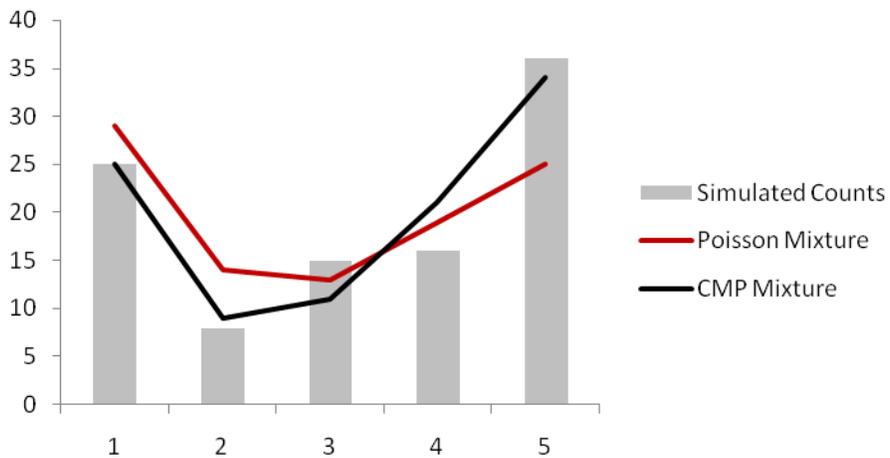

Figure 7: Simulated 5-point data and expected counts from Poisson and CMP mixture

## 3.3 Example 3: Bimodal distribution on 15-point scale

To further illustrate the ability of the CMP mixture to identify the two modes and adequately capture their frequency, as well as dips and overall shape, we further simulated two sets of 15-point scale data, with n=1,000 for each set. Table 4, Table 5 and Figure 8, Figure 9 present the simulated data, the fitted Poisson mixture and the fitted CMP mixture.

In the first example (Table 4 and Figure 8), both Poisson and CMP mixtures correctly identify the first mode (at 4), but the CMP estimates the corresponding peak much more accurately than the Poisson mixture. The second mode (at 15) is only identified correctly by the CMP mixture, whereas the Poisson mixture indicates a neighboring value (14) as the second mode. In terms of dips, the first lode (1) is identified by both models. However, for *lode$_2$=12* the Poisson estimate is far away at 8, while the CMP estimate is at the neighboring 11. Overall, the shape estimated by the CMP is dramatically closer to the data than the shape estimated by the Poisson mixture.

The second example (Table 5 and Figure 9) illustrates the dramatic under-estimation of a mode's peak magnitude using the Poisson mixture. In this example, while both Poisson and CMP mixtures reasonably capture the modes and lodes (with the CMP capturing them more accurately), they differ significantly in their estimate for the magnitude of the first peak. Such data shapes would not be uncommon in rating data.

Table 4: Simulated 15-point data and expected counts from Poisson and CMP mixtures

| Value | Simulated Counts | Poisson Mixture | CMP Mixture |
|---|---|---|---|
| 1 | 44 | 29 | 33 |
| 2 | 71 | 62 | 73 |
| 3 | 120 | 90 | 113 |
| 4 | 128 | 98 | 134 |
| 5 | 104 | 86 | 131 |
| 6 | 106 | 65 | 108 |
| 7 | 85 | 48 | 78 |
| 8 | 54 | 40 | 50 |
| 9 | 36 | 42 | 30 |
| 10 | 25 | 51 | 18 |
| 11 | 19 | 63 | 15 |
| 12 | 15 | 75 | 20 |
| 13 | 30 | 83 | 34 |
| 14 | 48 | 85 | 60 |

| | 15 | 115 | 83 | 103 |
|---|---|---|---|---|
| Estimates | | | | |
| $p$ | | 0.8 | 0.50 | 0.77 |
| $\lambda_1, \lambda_2$ | | 2,15 | 4.32,14.50 | 4.15,15.1 |
| $\nu_1, \nu_2$ | | 0.5,0.7 | | 0.9, 0.8 |
| First Mode | | 4 | 4 | 4 |
| Second Mode | | 15 | 14 | 15 |
| First Lode | | 1 | 1 | 1 |
| Second Lode | | 12 | 8 | 11 |
| Third Lode | | - | - | - |
| AIC | | | 5680 | 5210 |

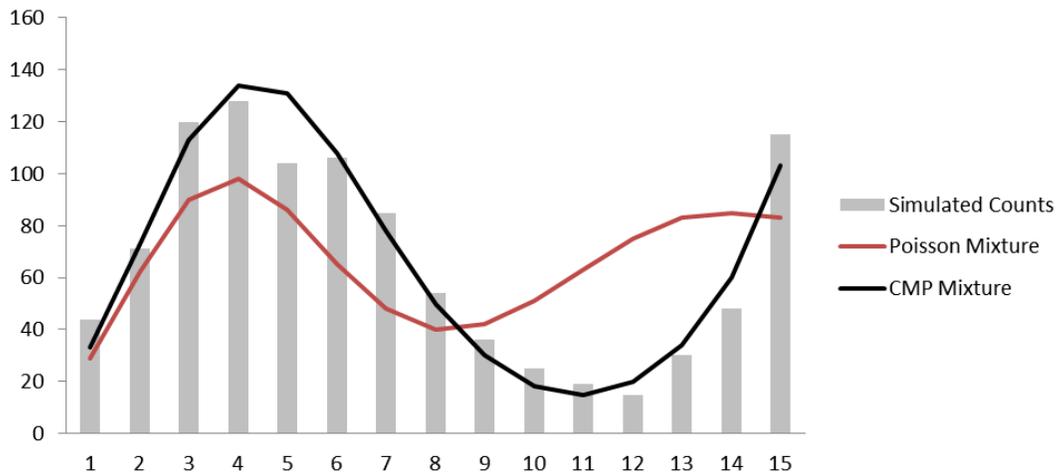

Figure 8: Simulated 15-point data (bars) and expected counts from Poisson and CMP mixtures

Table 5: Simulated 15-point data (n=1000), with estimated counts by Poisson and CMP mixtures

| Value | Simulated Counts | Poisson Mixture | CMP Mixture |
|---|---|---|---|
| 1 | 302 | 141 | 304 |
| 2 | 115 | 49 | 112 |
| 3 | 24 | 18 | 26 |
| 4 | 13 | 20 | 13 |
| 5 | 21 | 36 | 22 |
| 6 | 37 | 59 | 39 |
| 7 | 51 | 81 | 57 |
| 8 | 81 | 99 | 72 |
| 9 | 80 | 107 | 79 |
| 10 | 84 | 104 | 77 |
| 11 | 64 | 92 | 67 |
| 12 | 49 | 74 | 53 |
| **13** | 36 | 56 | 38 |
| **14** | 30 | 39 | 25 |
| **15** | 13 | 25 | 16 |
| Estimates |  |  |  |
| p | 0.4 | 0.20 | 0.44 |
| $\lambda_1, \lambda_2$ | 1,15 | 0.67,9.73 | 1.03,13.78 |
| $\nu_1, \nu_2$ | 1.5,1.2 |  | 1.5, 1.15 |
| First Mode | 1 | 1 | 1 |
| Second Mode | 10 | 9 | 9 |
| First Lode | 4 | 4 | 4 |
| Second Lode | 15 | 15 | 15 |
| Third Lode | - | - | - |
| AIC |  | 5050 | 4720 |

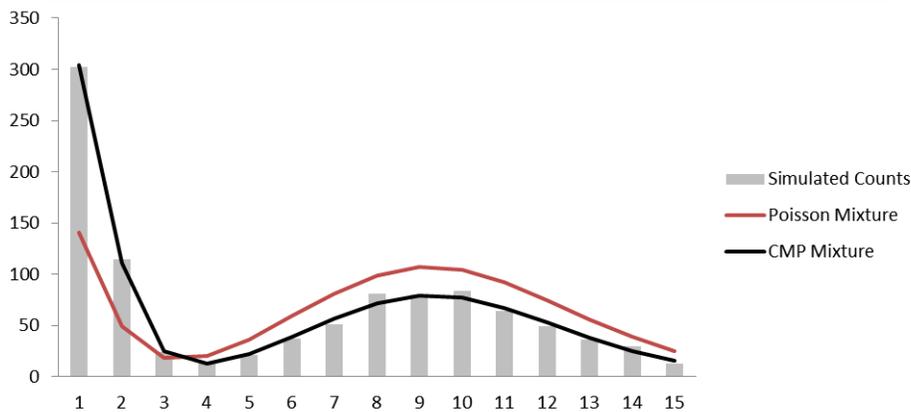

Figure 9: Simulated 15-point data (n=1000), with expected counts by Poisson and CMP mixtures

## 4  Application to Real Data

We now return to the three real-life examples presented in Section 1. In each case, we fit a CMP mixture, evaluate its fit, and compare it to a Poisson mixture.

### 4.1  Example 1: Market research survey question

Returning to the survey question regarding the presence of ice pieces in an ice-cream product (see Section 1.1), the results of fitting a CMP mixture and a Poisson mixture to the aggregated responses are shown in Table 6 and Figure 10. Although the fit of both models is not perfect, the CMP mixture better captures the overall shape in terms of the bimodal behavior and dip at the category "somewhat low".

Table 6: Observed and fitted models for ice-cream market research question

| Rating | Data | Poisson Mixture | CMP Mixture |
|---|---|---|---|
| ice absent | 39 | 31 | 36 |
| ice present somewhat low | 9 | 42 | 34 |
| neutral | 75 | 47 | 46 |
| ice present somewhat high | 52 | 45 | 47 |
| ice present very high | 24 | 35 | 36 |
| **Estimates** | | | |
| $p$ | | 0.1453 | 0.11 |
| $\lambda_1, \lambda_2$ | | **1.2, 3.9594** | 0.92, 4.98 |
| $\nu_1, \nu_2$ | | | 4.6, 1.2 |
| First Mode | Neutral | **neutral** | ice present somewhat high |
| Second Mode | ice absent | - | **Ice absent** |
| Dip Location | Ice present somewhat low | ice absent | **ice present somewhat low** |

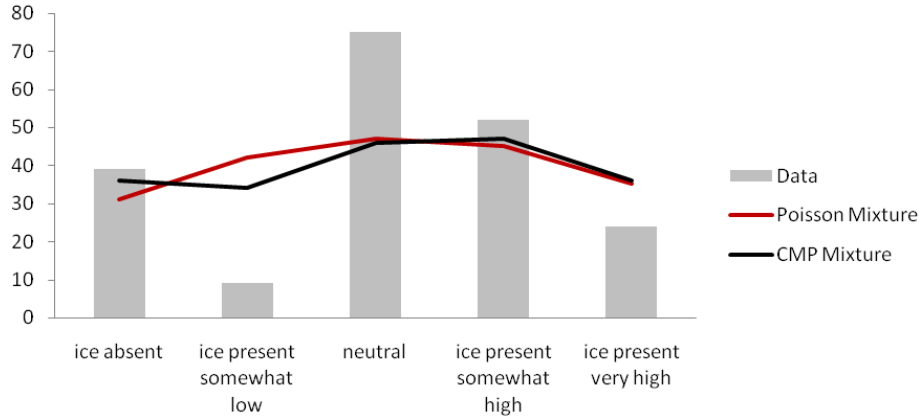

Figure 10: Observed and fitted counts for market research ice-cream question

## 4.2 Example 2: Heritage insurance competition

We return to the example from Section 1.2. The results of fitting a Poisson mixture and CMP mixture are shown in Table 7 and Figure 11. In this example, the two likelihood-based measures are very similar but the CMP fit is visibly much better. The CMP mixture correctly identifies the two modes and the magnitude of their frequencies. In contrast, the Poisson mixture not only misses the mode locations, but also the magnitude of the inaccuracy for those frequencies is quite high.

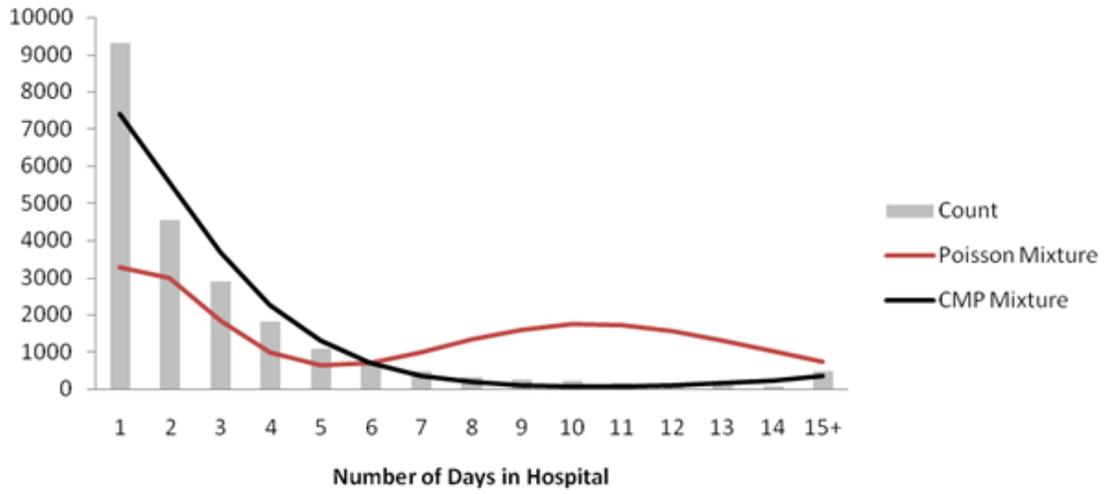

Figure 11: Observed and fitted Poisson and CMP mixture counts for Heritage Insurance Competition data

Table 7: Observed and fitted counts for Heritage Insurance Competition data

| # Days in hospital | Data | Poisson Mixture | CMP Mixture |
|---|---|---|---|
| 1 | 9299 | 3284 | 7410 |
| 2 | 4548 | 2994 | 5567 |
| 3 | 2882 | 1860 | 3704 |
| 4 | 1819 | 976 | 2260 |
| 5 | 1093 | 641 | 1290 |
| 6 | 660 | 713 | 698 |
| 7 | 474 | 994 | 361 |
| 8 | 316 | 1327 | 183 |
| 9 | 263 | 1600 | 96 |
| 10 | 209 | 1742 | 62 |
| 11 | 145 | 1725 | 62 |
| 12 | 135 | 1566 | 89 |
| 13 | 111 | 1313 | 142 |
| 14 | 65 | 1021 | 227 |
| 15+ | 479 | 742 | 347 |
| **Estimates** | | | |
| $p$ | | 0.4132 | 0.96 |
| $\lambda_1, \lambda_2$ | | 1.8156, 10.8937 | 0.93, 13.4 |
| $\nu_1, \nu_2$ | | | 0.3, 0.8 |
| **First Mode** | 1 | 1 | 1 |
| **Second Mode** | 15+ | 10 | 15+ |
| **Dip Location** | 14 | 5 | 10-11 |
| **AIC** | | 112006 | **85010** |

## 4.3 Example 3: Online ratings

Recall the Tripadvisor.com 5-point rating of Druk Hotel from Section 1.3. The results of fitting a Poisson mixture and CMP mixture are shown in Table 8 and Figure 12. A visual inspection shows that the CMP mixture outperforms the Poisson mixture in terms of capturing the overall shape of the distribution.

Table 8: Observed and fitted counts for Druk Hotel online ratings

| Rating | Data | Poisson Mixture | CMP Mixture |
|---|---|---|---|
| **Terrible** | 4 | 9 | 3 |
| **Poor** | 2 | 9 | 5 |
| **Average** | 10 | 9 | 8 |
| **very good** | 17 | 10 | 14 |
| **Excellent** | 17 | 13 | 19 |
| **Estimates** | | | |
| $p$ | | 0.22 | 0.09 |
| $\lambda_1, \lambda_2$ | | 1.58, 6.91 | 0.91, 5.23 |
| $\nu_1, \nu_2$ | | | 0.5, 0.8 |
| **First Mode** | very good, excellent | **excellent** | **excellent** |
| **Second Mode** | terrible | - | - |
| **Dip Location** | poor | terrible, poor, average | terrible |
| **AIC** | | 178.3156 | **171.1** |

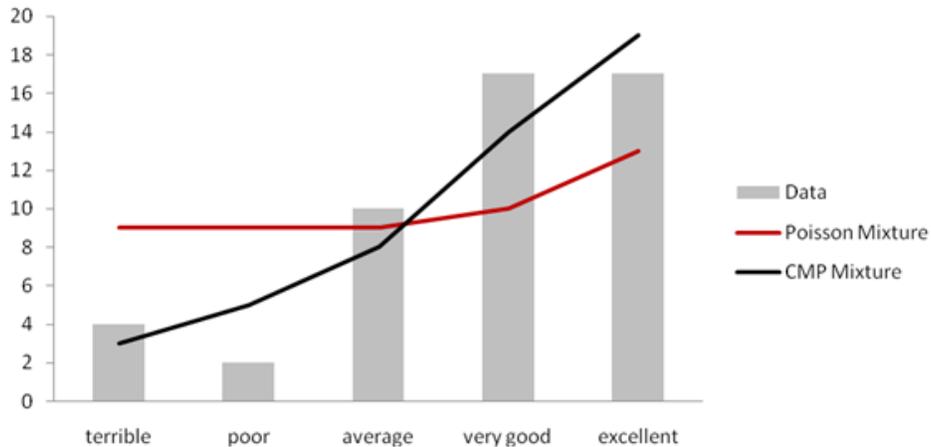

Figure 12: Observed and fitted Poisson and CMP mixture counts for Druk Hotel online rating example

In this example and in ratings applications in general, it is possible to flip the order of the values from low to high or from high to low. Here, we can re-order the ratings from "excellent" to "terrible". Next, we show the results of fitting Poisson and CMP mixtures to the flipped ratings (see Table 9 and Figure 13). It is interesting to note that for the CMP mixture the estimates slightly change, but the fitted counts remain unchanged. In contrast, for the Poisson mixture, flipping the order yields a slightly better fit in terms of shape.

Table 9: Poisson and CMP mixtures fitted to the flipped ratings (excellent to terrible)

| Rating | Data | Poisson Mixture | CMP Mixture |
|---|---|---|---|
| **Excellent** | 17 | 15 | 19 |
| **very good** | 17 | 14 | 14 |
| **Average** | 10 | 10 | 8 |
| **Poor** | 2 | 7 | 5 |
| **terrible** | 4 | 4 | 3 |
| **Estimates** | | | |
| $p$ | | 0.55 | 0.88 |
| $\lambda_1, \lambda_2$ | | 1.38, 3.38 | 1.03, 4.68 |
| $\nu_1, \nu_2$ | | | 0.6, 0.8 |
| **First Mode** | very good, excellent | **excellent** | **excellent** |
| **Second Mode** | terrible | - | - |
| **Dip Location** | poor | terrible, poor, average | terrible |
| **AIC** | | 206.8623 | **204.1** |

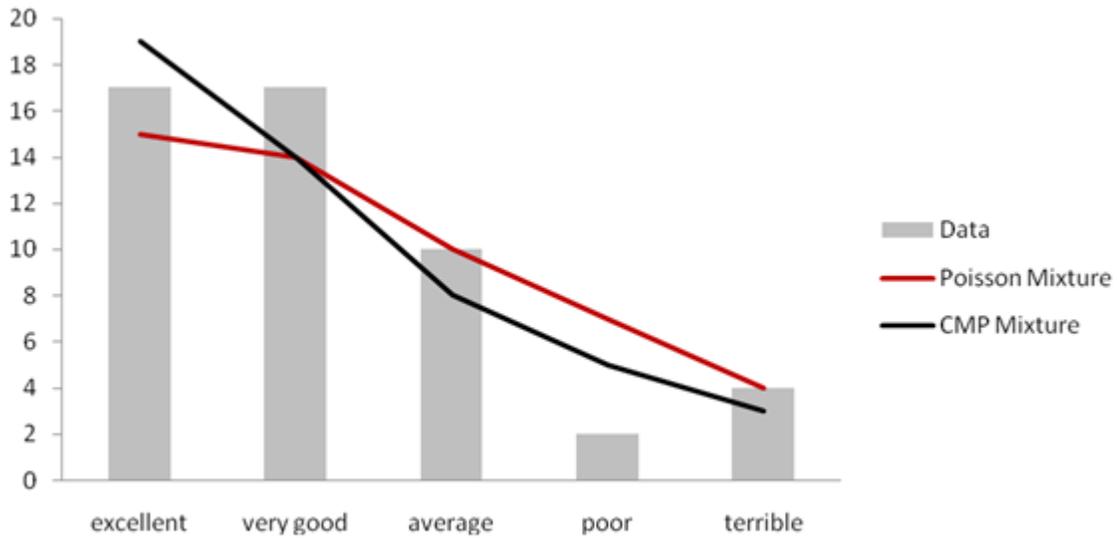

Figure 13: Poisson and CMP mixtures, fitted to the flipped ratings (excellent to terrible)

# 5   Discussion and Future Directions

Discrete data often exhibit bimodality that is difficult to model with standard distributions. A natural choice would be a mixture of two (or more) Poisson distributions. However due to the presence of under- or over-dispersion, often the Poisson mixture appears to be inadequate. The more general CMP distribution can capture under- or over-dispersion in the data. Therefore a mixture of CMP distributions (if necessary, properly truncated) may be appropriate to model such data.

The usual EM algorithm for fitting mixtures of distribution can be employed in this scenario. However, as the CMP distribution has an additional parameter (compared to the Poisson distribution), the maximization of the likelihood is nontrivial. In the absence of closed form solutions, iterative numerical algorithms are used for this purpose. An innovative two-step optimization with more than one possible initialization of the parameters has been suggested to ensure and speed up the convergence of the resulting algorithm. In our experiments, the proposed algorithm for fitting CMP mixture models takes less than two minutes even for very large datasets (such as the Example 6.2: Heritage Competition dataset). Further reduction in runtime may be possible by invoking more efficient optimization techniques.

An interesting property was observed while fitting the mixture of CMP distributions. If the ordering of the labels are reversed in case of (for example) consumer evaluation data, the fit appears to be very similar to the original one. This was not the case for the mixture of Poisson distributions. However this has to be more thoroughly investigated.

Though there is an inherent identifiability problem in the case of CMP mixture models, as there may be more than one combination of parameters of the underlying distributions yielding very similar shapes for the resulting mixtures, it does not cause any problem in terms of prediction. Rather it provides flexibility of choosing a model among several competing ones for improving predictive accuracy. Even for purposes of descriptive modeling, where we are interested in an approximation of the empirical distribution shape (location of peaks, etc.), the identifiability issue is not a challenge. It would only pose a challenge if the goal is identifying the underlying dispersion levels of the CMP distributions.

While Poisson and CMP distributions are designed for modeling count data, we note their usefulness in the context of bimodal discrete data that can include not only count data but also ordinal data such as ratings and rankings. Our illustrations show that using the CMP mixture can adequately capture the distribution of a sample from Likert-type scales and star ratings.

In our mixture scenario, observations are assumed to arise from a mixture distribution where it is not possible to identify which observation came from which original distribution ($CMP_1$ or $CMP_2$). Related work by Sellers and Shmueli (2012) uses a CMP regression formulation where predictor information is used to try and separate observations into dispersion groups and estimate the separate group-level dispersion. They show that mixing different dispersion levels can result in data with unexpected dispersion magnitude (e.g., mixing two under-dispersed CMPs can result in an apparent over-dispersed distribution). Our work differs from that work not only in looking at truncated CMPs, but also in the focus on predictive and descriptive modeling, where the goal is to find a parsimonious approximation for the observed empirical distribution.

This novel idea of CMP mixture modeling may also be extended to regression problems involving discrete bimodal data. For example, the Health Heritage example that we used comes from a larger contest for predicting length of stay at the hospital, where the data included many potential predictor variables. If the dependent variable shows bimodality, as in the case of the truncated "days in hospital" variable, the ordinary CMP regression might not be able to capture this feature. CMP mixture models may be very useful in this scenario. Another potential extension is to consider mixtures of more than two CMP distributions. Sellers and Shmueli (2010) considered CMP regression models for censored data. It would be interesting to explore the possibility of using a CMP mixture model in this context as well.